\documentclass[conference]{IEEEtran}
\IEEEoverridecommandlockouts

\usepackage{cite}
\usepackage{amsmath,amssymb,amsfonts}
\usepackage{algorithmic}
\usepackage{graphicx}
\usepackage{textcomp}
\usepackage{xcolor}
\usepackage{booktabs}  
\usepackage{siunitx}   
\usepackage{etoolbox}  
\usepackage{algorithm}
\usepackage{caption}
\usepackage{amsmath} 
\usepackage{multicol} 
\usepackage{multirow} 
\usepackage{graphicx}  
\usepackage{orcidlink}

\newrobustcmd{\B}{\bfseries} 
\def\BibTeX{{\rm B\kern-.05em{\sc i\kern-.025em b}\kern-.08em
    T\kern-.1667em\lower.7ex\hbox{E}\kern-.125emX}}
\begin{document}

\title{WhisQ: Cross-Modal Representation Learning for Text-to-Music MOS Prediction\\

% \thanks{Hokkaido Denshiki Co. LTD}
}

\author{%
  % 1st author: no department, ORCID + email
  \IEEEauthorblockN{Jakaria Islam Emon\orcidlink{0009-0002-3740-3818}}
  \IEEEauthorblockA{%
    Hokkaido Denshikiki Co., Ltd.\\
    Sapporo, Japan\\
    \href{mailto:emon_j@hdks.co.jp}{emon\_j@hdks.co.jp}
  }
  \and
  % 2nd author: has department
  \IEEEauthorblockN{Kazi Tamanna Alam}
  \IEEEauthorblockA{%
    Department of Computer Science and Engineering\\
    Barisal Information Technology College\\
    Barisal, Bangladesh\\
    \href{mailto:kazitamannat1@gmail.com}{kazitamannat1@gmail.com}
  }
  \and
  
  % 3rd author: no department, just company
  \IEEEauthorblockN{Md.\ Abu Salek\orcidlink{0009-0001-3747-1282}}
  \IEEEauthorblockA{%
    Hokkaido Denshikiki Co., Ltd.\\
    Sapporo, Japan\\
    \href{mailto:salek_a@hdks.co.jp}{salek\_a@hdks.co.jp}
    % or \href{https://orcid.org/YYYY-YYYY-YYYY-YYYY}{ORCID: YYYY-YYYY-YYYY-YYYY}
  }
}

\maketitle
\begin{abstract}
Mean Opinion Score (MOS) prediction for text-to-music systems requires evaluating both overall musical quality and text-prompt alignment. This paper introduces WhisQ, a multimodal architecture that addresses this dual-assessment challenge through sequence-level co-attention and optimal transport regularization. WhisQ employs the Whisper-Base pretrained model for temporal audio encoding and Qwen-3, a 0.6B Small Language Model (SLM), for text encoding, with both maintaining sequence structure for fine-grained cross-modal modeling. The architecture features specialized prediction pathways: OMQ is predicted from pooled audio embeddings, while TA leverages bidirectional sequence co-attention between audio and text. Sinkhorn optimal transport loss further enforce semantic alignment in the shared embedding space. On the MusicEval Track-1 dataset, WhisQ achieves substantial improvements over the baseline: 7\% improvement in Spearman correlation for OMQ and 14\% for TA. Ablation studies reveal that optimal transport regularization provides the largest performance gain (10\% SRCC improvement), demonstrating the importance of explicit cross-modal alignment for text-to-music evaluation.
\end{abstract}


\section{Introduction}
\label{sec:introduction} 

Recent advances in text-to-music (TTM) generation systems~\cite{agostinelli2023musiclm,liu2023audioldm,evans2024fast} have unlocked convincing, prompt conditioned music but have also created a pressing need for \emph{accurate, fast, and scalable} evaluation. The gold standard remains crowd-sourced \emph{mean opinion scores} (MOS), in which human listeners assign ratings on a five-point scale. Unfortunately, large-batch MOS collection is costly, time-consuming, and impractical for rapid iteration on ever-larger TTM models.

TTM evaluation is doubly challenging because it must assess not only the \textbf{overall musical quality (OMQ)} timbral fidelity, absence of artefacts, musicality but also the \textbf{textual alignment (TA)} between the generated audio and its prompt (e.g.\ “slow lo-fi waltz in 3/4 with pizzicato strings”). The AudioMOS Challenge formalises these two axes on the \textsc{MusicEval} dataset~\cite{MusicEval}, providing separate human labels for each.

Existing learned MOS predictors perform well for speech and generic audio~\cite{jiang2024sslmos,clap2023,Lo2019MOSNet}; however, they are not explicitly designed for the \emph{cross-modal semantic alignment} required by TTM tasks, and thus correlate poorly with TA.

To bridge this gap we introduce \textbf{WhisQ}, a fully automatic evaluator built on \emph{pretrained foundation models}:  
\textbf{(1) Whisper-Base}~\cite{radford2023robust}, an automatic-speech-recognition (ASR) model whose encoder provides strong  audio representations, and  
\textbf{(2) Qwen-3-0.6B} \cite{bai2023qwen}, a small language model (SLM) whose hidden states capture rich prompt semantics.  
WhisQ fuses these embeddings with sequence-level co-attention and enforces fine-grained correspondence through an \emph{optimal-transport regularization} term. 

\begin{figure}[t]
    \centering
    \includegraphics[width=1\linewidth]{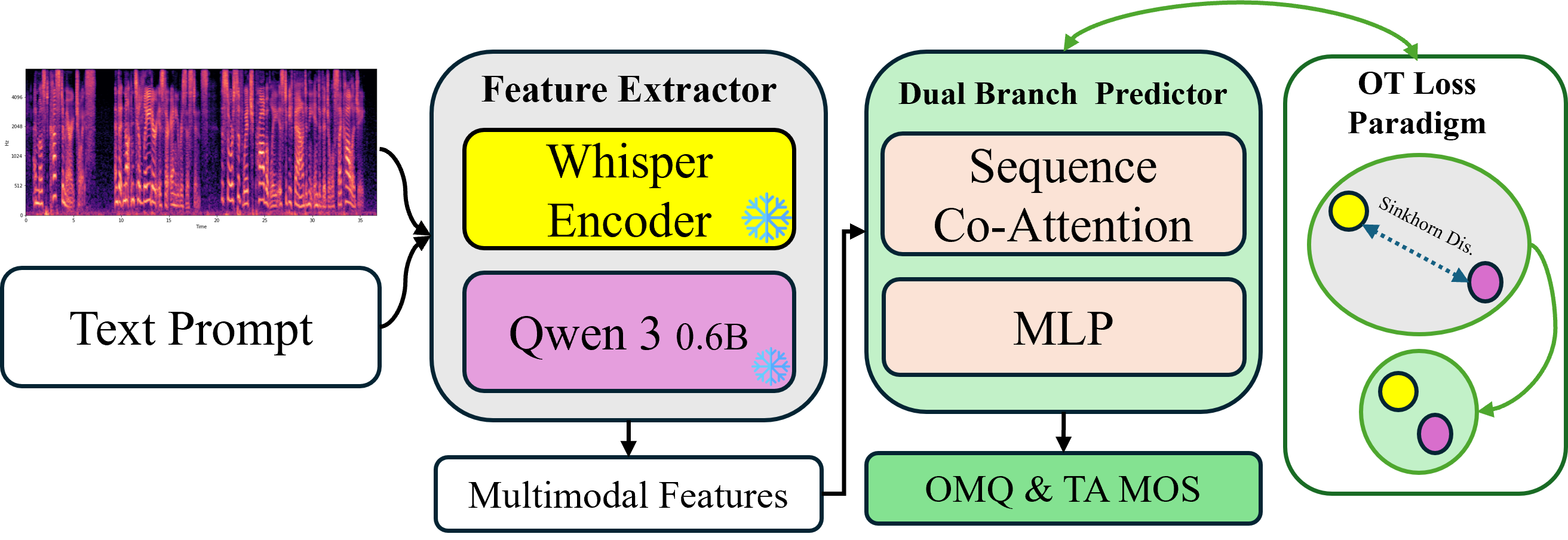}
    \caption{Overview of the proposed \textbf{WhisQ} architecture.  
    A log-mel spectrogram and its text prompt are encoded by pretrained speech and language models.  
    Sequence-level co-attention merges the embeddings; \textbf{two} lightweight MLP heads predict OMQ and TA MOS, while a Sinkhorn optimal transport loss tightens audio text alignment.}
    \label{fig:whisq_ot_architecture}
\end{figure}

The main contributions of this work are as follows:
\begin{itemize}
  \item \textbf{WhisQ}, a cross-modal representation learning framework for automatic MOS prediction in TTM.
  \item An optimal-transport regularizer that enforces fine-grained semantic alignment, yielding a 10\,\% boost in Spearman’s rank-correlation coefficient (SRCC) on the TA MOSover an ablated variant.
  \item State-of-the-art performance on \textsc{MusicEval} Track-1: +7\,\% SRCC on OMQ MOS and +14\,\% on TA MOS relative to the official baseline.
\end{itemize}

The rest of this paper is organized as follows: we detail our proposed architecture in Section~\ref{sec:proposed_method}, describe the experimental setup in Section~\ref{sec:experiments}, present the results in Section~\ref{sec:results}, and conclude with a discussion on future work in Section~\ref{sec:conclusion}.

\section{Proposed Method}
\label{sec:proposed_method}
This section details the \textbf{WhisQ} architecture: frozen foundation encoders, a 4-head sequence level co-attention module, and dual-branch head for MOS prediction as illustrated in Fig.~\ref{fig:whisq_ot_architecture}.

\subsection{Architecture}
\label{subsec:architecture}
Let $D_{\text{feat}} = 512$ be the primary feature dimension. Given an audio input $X_a$ and text prompt $X_t$, we extract sequence representations $\mathbf{H}_a \in \mathbb{R}^{T_a \times D_{\text{feat}}}$ and $\mathbf{H}_t \in \mathbb{R}^{T_t \times D_{\text{feat}}}$:
\begin{align}
\mathbf{H}_a &= \text{WhisperEncoder}(\text{LogMel}(X_a)) \label{eq:Ha_final} \\
\mathbf{H}_t &= \text{Proj}(\text{Qwen}(X_t)) \label{eq:Ht_final}
\end{align}
The Whisper-Base encoder provides $D_{\text{feat}}$-dimensional audio features from Log-Mel spectrograms. Outputs from the Qwen language model are processed by a linear projection layer, $\text{Proj}(\cdot)$, to yield $D_{\text{feat}}$-dimensional text features, ensuring alignment.  We employ sequence-level multi-head co-attention (MHA) to capture fine-grained temporal correspondences:
\begin{align}
\mathbf{H}_a^{\text{att}} &= \text{MHA}(\mathbf{H}_a, \mathbf{H}_t, \mathbf{H}_t) \in \mathbb{R}^{T_a \times D_{\text{feat}}} \label{eq:Ha_att_final} \\
\mathbf{H}_t^{\text{att}} &= \text{MHA}(\mathbf{H}_t, \mathbf{H}_a, \mathbf{H}_a) \in \mathbb{R}^{T_t \times D_{\text{feat}}} \label{eq:Ht_att_final}
\end{align}
The attended $D_{\text{feat}}$-dimensional sequences, $\mathbf{H}_a^{\text{att}}$ and $\mathbf{H}_t^{\text{att}}$, are then individually pooled by a function $\text{Pool}(\cdot)$ to $D_{\text{feat}}$-dimensional vectors. For TA prediction, these vectors are concatenated:
\begin{equation}
\mathbf{e}_{\text{seq}} = [\text{Pool}(\mathbf{H}_a^{\text{att}}); \text{Pool}(\mathbf{H}_t^{\text{att}})] \in \mathbb{R}^{2 \cdot D_{\text{feat}}} \label{eq:e_seq_final}
\end{equation}

For OMQ prediction, the initial audio features $\mathbf{H}_a$ are similarly pooled by $\text{Mean Pool}(\cdot)$ to obtain $\bar{\mathbf{H}}_a \in \mathbb{R}^{D_{\text{feat}}}$. Two distinct MLPs then predict the MOS scores:
\begin{align}
\hat{y}_{\text{OMQ}} &= \text{MLP}_{\text{OMQ}}(\bar{\mathbf{H}}_a) \label{eq:y_omq_final} \\
\hat{y}_{\text{TA}} &= \text{MLP}_{\text{TA}}(\mathbf{e}_{\text{seq}}) \label{eq:y_ta_final}
\end{align}
The MLP architectures are $\text{MLP}_{\text{OMQ}}: D_{\text{feat}} \to 256 \to 64 \to 1$ and $\text{MLP}_{\text{TA}}: (2 \cdot D_{\text{feat}}) \to 256 \to 64 \to 1$. Both utilize ReLU activations in their hidden layers.

\subsection{Training Objective}
\label{sec:training_objective}

The total training loss $\mathcal{L}$ combines a task-specific loss $\mathcal{L}_{\text{task}}$ for MOS prediction (OMQ and TA) with a Sinkhorn optimal transport loss $\mathcal{L}_{\text{OT}}$ \cite{an2022efficient} for cross-modal alignment, balanced by a hyperparameter $\lambda$:
\begin{equation}
\mathcal{L} = \mathcal{L}_{\text{task}} + \lambda \mathcal{L}_{\text{OT}}
\label{eq:total_loss}
\end{equation}

The task loss $\mathcal{L}_{\text{task}}$ addresses OMQ and TA. For TA predictions that utilize sequence co-attention, $\mathcal{L}_{\text{task}}$ is the average of Huber losses applied to the OMQ and sequence-level TA prediction heads:
\begin{equation}
\mathcal{L}_{\text{task}} = \frac{1}{2}\left[\mathcal{L}_{\text{Huber}}(\hat{y}_{\text{OMQ}}, y_{\text{OMQ}}^*) + \mathcal{L}_{\text{Huber}}(\hat{y}_{\text{TA}}^{\text{seq}}, y_{\text{TA}}^*)\right]
\label{eq:task_loss}
\end{equation}
where $\hat{y}$ denotes predicted scores (for OMQ and sequence-level TA, $\hat{y}_{\text{TA}}^{\text{seq}}$) and $y^*$ their corresponding ground-truth MOS labels. The Sinkhorn optimal transport loss $\mathcal{L}_{\text{OT}}$ enforces semantic alignment by minimizing the Sinkhorn distance (which approximates optimal transport) between sequences of audio embeddings $\mathbf{H}_a^{(i)}$ and text embeddings $\mathbf{H}_t^{(i)}$. This is averaged over a batch of size $B$:
\begin{equation}
\mathcal{L}_{\text{OT}} = \frac{1}{B}\sum_{i=1}^{B} \text{Sinkhorn}(\mathbf{H}_a^{(i)}, \mathbf{H}_t^{(i)})
\label{eq:ot_loss}
\end{equation}
Minimizing $\mathcal{L}_{\text{OT}}$ encourages the distributions of the audio encoder outputs $\mathbf{H}_a$ and text encoder outputs $\mathbf{H}_t$ to become more similar in the shared embedding space.

\section{Experiments}
\label{sec:experiments}

\subsection{Dataset}
\label{ssec:dataset}
We conduct our experiments on Track 1 of the AudioMOS Challenge, which uses the MusicEval dataset \cite{MusicEval}. It comprises 2,748 music clips generated by 31 different text-to-music models in response to 384 unique text prompts. Each music clip was evaluated by 5 human raters (14 music experts in total) across two dimensions: 
\begin{itemize}
    \item \textbf{OMQ MOS:} Reflecting the overall musical impression.
    \item \textbf{TA MOS:} Assessing the consistency of the generated music with the given text prompt.
\end{itemize}
Our development and ablation studies are performed on the provided training and validation sets.

\subsection{Implementation Details}
\label{ssec:implementation_details}
WhisQ is implemented in PyTorch and trained on a single NVIDIA RTX 3060 with automatic mixed precision and gradient clipping. The pretrained encoders (Whisper-Base for audio and Qwen3 0.6B for text) are frozen, while the projection layer, 4-head sequence-level co-attention block, and two MLP prediction heads are updated. Audio is resampled to 16 kHz and padded or truncated to 3000 log-mel frames. Remaining optimization details and hyper-parameter values are given in Table~\ref{tab:hyperparameters}.
The proposed WhisQ model has 619.45M total parameters and requires 32.54 Giga MACs (GMACs) for inference. However, since we utilize frozen backbones, the number of task-specific trainable parameters (from the attention module and MLP heads) is only 2.66M.

\begin{table}[htbp]
\centering
\caption{Optimized hyperparameters for WhisQ.}
\label{tab:hyperparameters}
\begin{tabular}{@{}ll@{}}
\toprule
Hyperparameter         & Value                       \\
\midrule
Learning Rate& \num{7.307e-4}              \\
Optimizer & SGD                         \\
Momentum& \num{0.7435}                \\
Batch Size& \num{128}                     \\
Epochs & \num{148}  \\
Loss Function& Huber                       \\
OT Method & Sinkhorn ($p$=2, blur=0.05)\\
OT Weight ($\lambda$)& \num{4.057e-5}             \\
Feature Dimension & \num{1280} \\
\bottomrule
\end{tabular}
\end{table}

\begin{table*}[htbp]
\centering
\caption{Comprehensive evaluation results on MusicEval Track-1 validation set. The table shows baseline comparison followed by systematic ablation studies examining attention mechanisms, OT alignment, and pretrained backbones.}
\label{tab:ablation_combined}
\resizebox{\textwidth}{!}{%
\begin{tabular}{@{}l@{\hspace{0.5em}}cccc@{\hspace{0.8em}}cccc@{\hspace{0.8em}}cccc@{\hspace{0.8em}}cccc@{}}
\toprule
& \multicolumn{8}{c}{Overall Quality (OQM)} & \multicolumn{8}{c}{Textual Alignment (TA)} \\
\cmidrule(lr){2-9} \cmidrule(lr){10-17}
& \multicolumn{4}{c}{Utterance-level} & \multicolumn{4}{c}{System-level} & \multicolumn{4}{c}{Utterance-level} & \multicolumn{4}{c}{System-level} \\
\cmidrule(lr){2-5} \cmidrule(lr){6-9} \cmidrule(lr){10-13} \cmidrule(lr){14-17}
Configuration & MSE↓ & LCC↑ & SRCC↑ & KTAU↑ & MSE↓ & LCC↑ & SRCC↑ & KTAU↑ & MSE↓ & LCC↑ & SRCC↑ & KTAU↑ & MSE↓ & LCC↑ & SRCC↑ & KTAU↑ \\
\midrule
\multicolumn{17}{l}{\textit{Baseline Comparison}} \\
\midrule
\textbf{WhisQ w/ OT (Proposed)} & \textbf{0.3584} & \textbf{0.7523} & \textbf{0.7585}& \textbf{0.5891}& \textbf{0.1095} & \textbf{0.9044}& \textbf{0.8813}& \textbf{0.7143}& \textbf{0.4735} & \textbf{0.6176} & \textbf{0.6109} & \textbf{0.4474} & \textbf{0.0773} & \textbf{0.8721} & \textbf{0.8695} & \textbf{0.6749} \\
WhisQ w/o OT  & 0.5705& 0.6807& 0.6941& 0.5165& 0.3118& 0.7986& 0.8079& 0.6108& 0.5688& 0.4939& 0.4895& 0.3481& 0.1442& 0.7613& 0.7828& 0.5961\\
Official Baseline & 0.6175 & 0.6908 & 0.6881 & 0.5143 & 0.3863 & 0.8016 & 0.7764 & 0.5862 & 0.5936 & 0.5803 & 0.5425 & 0.3933 & 0.2322 & 0.7461 & 0.7202 & 0.5074 \\
\midrule
\multicolumn{17}{l}{\textit{Attention Mechanism Ablations}} \\
\midrule
Vanilla Co-Attention w/ OT& 0.4249 & 0.7156 & 0.7230 & 0.5411 & 0.1129 & 0.8884 & 0.8757 & 0.6896 & 0.5123 & 0.5934 & 0.5876 & 0.4287 & 0.0896 & 0.8456 & 0.8321 & 0.6234 \\
Cross-Attention w/ OT & 0.4601 & 0.6741 & 0.6881 & 0.5038 & 0.1193 & 0.8721 & 0.8606 & 0.6700 & 0.5567 & 0.5645 & 0.5523 & 0.3987 & 0.0987 & 0.8234 & 0.8156 & 0.5987 \\
MLP Only& 0.4857 & 0.6487 & 0.6695 & 0.4878 & 0.1764 & 0.8582 & 0.8507 & 0.6798 & 0.5789 & 0.5456 & 0.5234 & 0.3856 & 0.1234 & 0.7987 & 0.7865 & 0.5678 \\
\midrule
\multicolumn{17}{l}{\textit{Pretrained Backbone Ablations (with Sequence Co-Attention + OT + Huber loss)}} \\
\midrule
Wav2Vec2 + ModernBERT & 0.9766 & -0.1206 & -0.1540 & -0.1044 & 0.5860 & -0.3033 & -0.3005 & -0.1921 & 0.7778 & -0.0291 &  0.0073 &  0.0061 & 0.3215 & -0.2869 & -0.0773 & -0.0690 \\

\bottomrule
\end{tabular}
}
\end{table*}

\section{Results and Analysis}
\label{sec:results}

We evaluate WhisQ on the MusicEval Track-1 validation set, reporting utterance-level (U) and system-level (S) metrics: Mean Squared Error (MSE$\downarrow$), Pearson Correlation (LCC$\uparrow$), Spearman Correlation (SRCC$\uparrow$), and Kendall's Tau (KTAU$\uparrow$). Table~\ref{tab:ablation_combined} presents comprehensive results, including baseline comparisons and systematic ablation studies.

\subsection{Performance Analysis}
\label{subsec:performance_analysis}

The performance of our proposed model, \textbf{WhisQ w/ OT}, is detailed in Table~\ref{tab:ablation_combined}, showcasing its effectiveness against the baseline and various ablated configurations.

\textbf{Baseline Comparison:}
WhisQ w/ OT (Proposed) significantly outperforms the Official Baseline across all evaluated metrics. For OQ) at the utterance-level, WhisQ achieves an SRCC of 0.7585 (compared to 0.6881 for the baseline) and reduces MSE to 0.3584 (from 0.6175). System-level OQM SRCC is also notably higher at 0.8813 (vs. 0.7764). Similar substantial gains are observed for TA, where utterance-level SRCC improves to 0.6109 (from 0.5425) and system-level SRCC reaches 0.8695 (vs. 0.7202 for the baseline).

\textbf{Impact of Optimal Transport (OT):}
The inclusion of Optimal Transport regularization proves critical for performance. Comparing \textbf{WhisQ w/ OT (Proposed)} to its variant \textbf{WhisQ w/o OT} reveals a clear benefit. For instance, removing OT causes the utterance-level TA SRCC to drop from 0.6109 to 0.4895. These results highlight the crucial role of OT in fostering effective cross-modal alignment.

\textbf{Attention Mechanisms:}
Our proposed sequence co-attention architecture within \textbf{WhisQ w/ OT (Proposed)} demonstrates the strongest performance among the evaluated attention variants (e.g., OQM U-SRCC 0.7585; TA S-SRCC 0.8695). Other configurations, such as \textbf{Vanilla Co-Attention w/ OT} (OQM U-SRCC 0.7230) and \textbf{Cross-Attention w/ OT} (OQM U-SRCC 0.6881), yield lower scores. The \textbf{MLP Only} configuration, which eschews sophisticated attention, performs the poorest among these (OQM U-SRCC 0.6695), underscoring the necessity of advanced attention mechanisms for this complex cross-modal task.

\textbf{Pretrained Backbones:}
Our proposed WhisQ substantially outperforms the alternative \textbf{Wav2Vec2 + ModernBERT} backbone. As shown in Table~\ref{tab:ablation_combined}, WhisQ achieves an OQM utterance-level SRCC of 0.7585, while the Wav2Vec2~\cite{baevski2020wav2vec} combined with ModernBERT~\cite{warner2024smarter} results in a negative SRCC of -0.1540 for the same metric. This difference underscores the superior suitability of the Whisper and Qwen 3 combination for encoding the audio and textual modalities effectively in the TTM evaluation context.

\section{Conclusion and Future Work}
\label{sec:conclusion}

In this paper, we introduced WhisQ, a novel system for automatic Mean Opinion Score prediction. By leveraging sequence-level co-attention and optimal transport regularization, WhisQ effectively models fine-grained audio-text correspondences and learns semantically aligned cross-modal representations. On the MusicEval Track-1 benchmark, WhisQ demonstrated significant improvements over the baseline, notably increasing the Spearman's rank-correlation coefficient (SRCC) by 7\% for OMQ and 14\% for TA. 

Despite these promising results, we acknowledge certain limitations in the current iteration of WhisQ. The primary concern is its computational footprint. 
Additionally, our approach currently relies on frozen backbones; while these provide strong initial representations, this strategy prevents end-to-end fine-tuning that could potentially unlock further performance gains.

Building on these findings and addressing current limitations, future research offers several promising avenues. Exploration of alternative alignment objectives, such as different optimal transport variants (e.g., Gromov-Wasserstein) or contrastive learning approaches, could further enhance cross-modal alignment, particularly for diverse musical styles. Investigating WhisQ's generalization is also important, including its evaluation on a broader range of audio quality datasets and its adaptation to related tasks like text-to-speech or general text-to-audio quality assessment. Finally, improving model interpretability for instance, by visualizing the learned cross-modal alignments can provide crucial insights into its music-text correspondence mechanisms, thereby informing future model refinements and the development of more effective TTM systems.

\nocite{*}
\bibliographystyle{IEEEtran}
\bibliography{refs}

\begin{thebibliography}{10}
\providecommand{\url}[1]{#1}
\csname url@samestyle\endcsname
\providecommand{\newblock}{\relax}
\providecommand{\bibinfo}[2]{#2}
\providecommand{\BIBentrySTDinterwordspacing}{\spaceskip=0pt\relax}
\providecommand{\BIBentryALTinterwordstretchfactor}{4}
\providecommand{\BIBentryALTinterwordspacing}{\spaceskip=\fontdimen2\font plus
\BIBentryALTinterwordstretchfactor\fontdimen3\font minus \fontdimen4\font\relax}
\providecommand{\BIBforeignlanguage}[2]{{%
\expandafter\ifx\csname l@#1\endcsname\relax
\typeout{** WARNING: IEEEtran.bst: No hyphenation pattern has been}%
\typeout{** loaded for the language `#1'. Using the pattern for}%
\typeout{** the default language instead.}%
\else
\language=\csname l@#1\endcsname
\fi
#2}}
\providecommand{\BIBdecl}{\relax}
\BIBdecl

\bibitem{agostinelli2023musiclm}
A.~Agostinelli, T.~I. Denk, Z.~Borsos, J.~Engel, M.~Verzetti, A.~Caillon, Q.~Huang, A.~Jansen, A.~Roberts, M.~Tagliasacchi \emph{et~al.}, ``Musiclm: Generating music from text,'' \emph{arXiv preprint arXiv:2301.11325}, 2023.

\bibitem{liu2023audioldm}
H.~Liu, Z.~Chen, Y.~Yuan, X.~Mei, X.~Liu, D.~Mandic, W.~Wang, and M.~D. Plumbley, ``Audioldm: Text-to-audio generation with latent diffusion models,'' \emph{arXiv preprint arXiv:2301.12503}, 2023.

\bibitem{evans2024fast}
Z.~Evans, C.~Carr, J.~Taylor, S.~H. Hawley, and J.~Pons, ``Fast timing-conditioned latent audio diffusion,'' in \emph{Forty-first International Conference on Machine Learning}, 2024.

\bibitem{MusicEval}
C.~Liu, H.~Wang, J.~Zhao, S.~Zhao, H.~Bu, X.~Xu, J.~Zhou, H.~Sun, and Y.~Qin, ``Musiceval: A generative music dataset with expert ratings for automatic text-to-music evaluation,'' 04 2025, pp. 1--5.

\bibitem{jiang2024sslmos}
Z.~Jiang, X.~Li, and P.~Lu, ``{SSL-MOS: A Self-Supervised Learning Based Approach with A Transformer Target Model For MOS Prediction},'' in \emph{ICASSP 2024 - 2024 IEEE International Conference on Acoustics, Speech and Signal Processing (ICASSP)}.\hskip 1em plus 0.5em minus 0.4em\relax IEEE, 2024, pp. 1061--1065, relevant for \textbackslash{}cite\{sslfinetune2024\}.

\bibitem{clap2023}
Y.~Wu, K.~Chen, T.~Zhang, Y.~Hui, T.~Berg-Kirkpatrick, and S.~Dubnov, ``Large-scale contrastive language-audio pretraining with feature fusion and keyword-to-caption augmentation,'' in \emph{ICASSP 2023 - 2023 IEEE International Conference on Acoustics, Speech and Signal Processing (ICASSP)}, 2023, pp. 1--5.

\bibitem{Lo2019MOSNet}
C.-C. Lo, S.-W. Fu, W.-C. Huang, H.-M. Wang, T.~Toda, and Y.~Tsao, ``{MOSNet}: Deep learning based objective assessment for voice conversion,'' in \emph{Interspeech 2019}, 2019, pp. 1133--1137.

\bibitem{radford2023robust}
A.~Radford, J.~W. Kim, T.~Xu, G.~Brockman, C.~McLeavey, and I.~Sutskever, ``Robust speech recognition via large-scale weak supervision,'' in \emph{International conference on machine learning}.\hskip 1em plus 0.5em minus 0.4em\relax PMLR, 2023, pp. 28\,492--28\,518.

\bibitem{bai2023qwen}
J.~Bai, S.~Bai, Y.~Chu, Z.~Cui, K.~Dang, X.~Deng, Y.~Fan, W.~Ge, Y.~Han, F.~Huang \emph{et~al.}, ``Qwen technical report,'' \emph{arXiv preprint arXiv:2309.16609}, 2023.

\bibitem{an2022efficient}
D.~An, N.~Lei, X.~Xu, and X.~Gu, ``Efficient optimal transport algorithm by accelerated gradient descent,'' in \emph{Proceedings of the AAAI Conference on Artificial Intelligence}, vol.~36, no.~9, 2022, pp. 10\,119--10\,128.

\bibitem{baevski2020wav2vec}
A.~Baevski, Y.~Zhou, A.~Mohamed, and M.~Auli, ``wav2vec 2.0: A framework for self-supervised learning of speech representations,'' \emph{Advances in neural information processing systems}, vol.~33, pp. 12\,449--12\,460, 2020.

\bibitem{warner2024smarter}
B.~Warner, A.~Chaffin, B.~Clavi{\'e}, O.~Weller, O.~Hallstr{\"o}m, S.~Taghadouini, A.~Gallagher, R.~Biswas, F.~Ladhak, T.~Aarsen \emph{et~al.}, ``Smarter, better, faster, longer: A modern bidirectional encoder for fast, memory efficient, and long context finetuning and inference,'' \emph{arXiv preprint arXiv:2412.13663}, 2024.

\bibitem{copet_simple_2023}
J.~Copet \emph{et~al.}, ``Simple and controllable music generation,'' in \emph{Advances in Neural Information Processing Systems}, 2023.

\bibitem{schneider_mousai_2024}
F.~Schneider, O.~Kamal, Z.~Jin, and B.~Schölkopf, ``Moûsai: Efficient text-to-music diffusion models,'' in \emph{Proceedings of the 62nd Annual Meeting of the Association for Computational Linguistics}, 2024.

\bibitem{zhang_musicmagus_2024}
Y.~Zhang \emph{et~al.}, ``Musicmagus: Zero-shot text-to-music editing via diffusion models,'' in \emph{Proceedings of the 33rd International Joint Conference on Artificial Intelligence}, 2024.

\bibitem{lelan_flowmatch_2024}
G.~Le~Lan \emph{et~al.}, ``High fidelity text-guided music generation and editing via single-stage flow matching,'' \emph{arXiv preprint arXiv:2407.03648}, 2024.

\bibitem{chen_musicldm_2023}
K.~Chen, Y.~Wu, H.~Liu, M.~Nezhurina, T.~Berg-Kirkpatrick, and S.~Dubnov, ``Musicldm: Enhancing novelty in text-to-music generation using beat-synchronous mixup strategies,'' \emph{arXiv preprint arXiv:2308.01546}, 2023.

\bibitem{wang_mospc_2023}
K.~Wang, Y.~Zhao, Q.~Dong, T.~Ko, and M.~Wang, ``Mospc: Mos prediction based on pairwise comparison,'' in \emph{Proceedings of the 61st Annual Meeting of the Association for Computational Linguistics}, 2023.

\bibitem{do_resource_2023}
P.~Do, M.~Coler, J.~Dijkstra, and E.~Klabbers, ``Resource-efficient fine-tuning strategies for automatic mos prediction in text-to-speech for low-resource languages,'' in \emph{Proceedings of Interspeech 2023}, 2023.

\bibitem{kunesova_voicemos_2024}
M.~Kunešová, J.~Lehečka, J.~Michálek, J.~Matoušek, and J.~Švec, ``Zero-shot out-of-domain is no joke: Lessons learned in the voicemos 2023 challenge,'' in \emph{Proceedings of Interspeech 2024}, 2024.

\bibitem{lian_apgmos_2025}
Z.~Lian, L.~Wang, and H.~Huang, ``Apg-mos: Auditory perception guided-mos predictor for synthetic speech,'' \emph{arXiv preprint arXiv:2504.20447}, 2025.

\bibitem{rho_lavcap_2025}
K.~Rho, H.~Lee, V.~Iverson, and J.~S. Chung, ``Lavcap: Llm-based audio-visual captioning using optimal transport,'' \emph{arXiv preprint arXiv:2501.09291}, 2025.

\bibitem{wang_usam_2025}
Z.~Wang, X.~Xia, X.~Zhu, and L.~Xie, ``U-sam: An audio language model for unified speech, audio, and music understanding,'' \emph{arXiv preprint arXiv:2505.13880}, 2025.

\bibitem{zhou_cmot_2023}
Y.~Zhou, Q.~Fang, and Y.~Feng, ``Cmot: Cross-modal mixup via optimal transport for speech translation,'' in \emph{Proceedings of the 61st Annual Meeting of the Association for Computational Linguistics}, 2023.

\end{thebibliography}

\end{document}